\documentclass[12pt]{article}
\usepackage{amssymb,epsf}
\def\lsim{\mathrel{\rlap {\raise.5ex\hbox{$ < $}}
{\lower.5ex\hbox{$\sim$}}}}
\def\gsim{\mathrel{\rlap {\raise.5ex\hbox{$ > $}}
{\lower.5ex\hbox{$\sim$}}}} 
\usepackage{epsf}
\def\sqr#1#2{{\vcenter{\vbox{\hrule height.#2pt

        \hbox{\vrule width.#2pt height#1pt \kern#1pt

           \vrule width.#2pt}

        \hrule height.#2pt}}}}

\def\lsim{{\displaystyle
{{\raise-8pt\hbox{$ <$}}
\atop{\raise5pt\hbox{$\sim$}}}}}
\def\gsim{{\displaystyle
{{\raise-8pt\hbox{$ >$}}
\atop{\raise5pt\hbox{$\sim$}}}}}
\def\slsim{{\displaystyle
{{\raise-8pt\hbox{$\scriptstyle <$}}
\atop{\raise5pt\hbox{$\scriptstyle \sim$}}}}}
\def\sgsim{{\displaystyle
{{\raise-8pt\hbox{$\scriptstyle  >$}}
\atop{\raise5pt\hbox{$\scriptstyle \sim$}}}}}
\newskip\humongous \humongous=0pt plus 1000pt minus 1000pt

\newcommand{\sumpf}[0]{\sum_{(H^{\rm f},G^{\rm f})}\! \! \! \!
{\raise
4pt
\hbox{$'$}}\,}

\newcommand{\sump}[0]{\sum_{(H,G)}\! \! {\raise 4pt \hbox{$'$}}\,}

\def\bs{\begin{subequations}}
\def\es{\end{subequations}}
\catcode`\@=11
\newcount\hour
\newcount\minute
\newtoks\amorpm
\hour=\time\divide\hour by60
\minute=\time{\multiply\hour by60 \global\advance\minute by-\hour}
\edef\standardtime{{\ifnum\hour<12 \global\amorpm={am}%
        \else\global\amorpm={pm}\advance\hour by-12 \fi
        \ifnum\hour=0 \hour=12 \fi
        \number\hour:\ifnum\minute<10 0\fi\number\minute\the\amorpm}}
\edef\militarytime{\number\hour:\ifnum\minute<10 0\fi\number\minute}
\def\draftlabel#1{{\@bsphack\if@filesw {\let\thepage\relax
   \xdef\@gtempa{\write\@auxout{\string
      \newlabel{#1}{{\@currentlabel}{\thepage}}}}}\@gtempa
   \if@nobreak \ifvmode\nobreak\fi\fi\fi\@esphack}
        \gdef\@eqnlabel{#1}}
\def\@eqnlabel{}
\def\@vacuum{}
\def\draftmarginnote#1{\marginpar{\raggedright\scriptsize\tt#1}}
\def\draft{\oddsidemargin -.2truein
        \def\@oddfoot{\sl preliminary draft \hfil
        \rm\thepage\hfil\sl\today\quad\militarytime}
        \let\@evenfoot\@oddfoot \overfullrule 3pt
        \let\label=\draftlabel
        \let\marginnote=\draftmarginnote
   \def\@eqnnum{(\theequation)\rlap{\kern\marginparsep\tt\@eqnlabel}%
\global\let\@eqnlabel\@vacuum}  }
\def\subequations{\refstepcounter{equation}%
  \edef\@savedequation{\the\c@equation}%
  \@stequation=\expandafter{\theequation}%   %only want \theequation
  \edef\@savedtheequation{\the\@stequation}% % expanded once
  \edef\oldtheequation{\theequation}%
  \setcounter{equation}{0}%
  \def\theequation{\oldtheequation\alph{equation}}}

\def\endsubequations{\setcounter{equation}{\@savedequation}%
  \@stequation=\expandafter{\@savedtheequation}%
  \edef\theequation{\the\@stequation}\global\@ignoretrue
  \vspace*{-12pt} \\}
\def\bs{\begin{subequations}}
\def\es{\end{subequations}}
\relax
\def\thefootnote{\fnsymbol{footnote}}
\def\be{\begin{equation}}
\def\ee{\end{equation}}
\def\ba{\begin{eqnarray}}
\def\ea{\end{eqnarray}}

\def\ee{\end{equation}}
\def\bea{\begin{eqnarray}}
\def\eea{\end{eqnarray}}
\def\nn{\nonumber}

\newcommand{\uarrw}[0]{\mathrel{
{\raise.5ex\vbox{\hrule width 1cm}\hskip-6pt\rightarrow}}}
\def\thebibliography#1{%
\vskip 0.5cm \centerline{\bf References}
\list{%
[\arabic{enumi}]}{\settowidth\labelwidth{[#1]}
\leftmargin\labelwidth
\advance\leftmargin\labelsep
\usecounter{enumi}}
\def\newblock{\hskip .11em plus .33em minus .07em}
\sloppy\clubpenalty4000\widowpenalty4000
\sfcode`\.=1000\relax}

\renewcommand{\theequation}{\arabic{section}.\arabic{equation}}

\renewcommand{\section}{\setcounter{equation}{0}\@startsection%
{section}{1}{0mm}{-\baselineskip}{0.5\baselineskip}%
{\normalfont\normalsize\bfseries}}

\renewcommand{\subsection}{\@startsection%
{subsection}{2}{0mm}{-\baselineskip}{0.5\baselineskip}%
{\normalfont\normalsize\slshape}}

\renewcommand{\subsubsection}{\@startsection%
{subsubsection}{2}{0mm}{-\baselineskip}{0.5\baselineskip}%
{\normalfont\normalsize\slshape}}

\begin{document}
%%
%\special{!userdict begin /bop-hook{gsave 200 30 translate
%65 rotate /Times-Roman findfont 216 scalefont setfont
%0 0 moveto 0.85 setgray (\jobname) show grestore}def end}
% 
\renewcommand{\theequation}{\arabic{section}.\arabic{equation}}
\begin{titlepage}
\begin{flushright}
%hep-th/yymmnnn
\end{flushright}
\begin{centering}
\vspace{2,5cm}
\boldmath
{ 
\bf \large A note on the phases of natural evolution
}
\\
\unboldmath
\vspace{1.9 cm}
{\bf Andrea Gregori}$^{\dagger}$ \\
\medskip
\vspace{.4in}

\vspace{2.2cm}
{\bf Abstract}\\
\vspace{.2in}
\end{centering}
The natural evolution of life seems to proceed through steps characterized
by phases of relatively rapid changes, followed by longer, more stable periods.
In the light of the string-theory derived physical scenario proposed
in \cite{spi}, we discuss how this behaviour can be related to
a sequence of resonances of the energy of natural sources of radiation
and absorption energies of the DNA, responsible for mutagenesis.
In a scenario in which these energy scales run independently as functions of  
the age of the Universe, the conditions for evolutionary mutagenesis
are satisfied only at discrete points of the time axis, and for a short 
period, corresponding to the width of the resonance. We consider in particular
the evolution of the primates through subsequent steps of increasing
cranio-facial contraction, and the great Eras of life
(Paleozoic, Mesozoic, Cenozoic), showing that the transitions occur at 
the predicted times of resonance.
  
\vspace{4cm}

\hrule width 6.7cm
\noindent
$^{\dagger}$e-mail: agregori@libero.it

\end{titlepage}
\newpage
\setcounter{footnote}{0}
\renewcommand{\thefootnote}{\arabic{footnote}}

\vspace{1.5cm}

Paleontological observations seem to indicate that the evolution of life 
would not take place ``progressively'', but would be characterized
by relatively short periods of ``sudden'' mutation, separated 
by longer, more or less ``stable'' periods. For instance, 
it has been observed that the species of hominids,
from the primates to the homo sapiens, is characterized by an evolution
toward an increasing cranio-facial contraction, which makes possible an
expansion of the volume of the brain, 
and appears to take place at specific periods in which a 
big step forward is made, followed by longer periods in which this kind of 
mutagenesis seems to be ``at rest'' \cite{malasse1}.  
This progressing through ``steps''
seems in some way to call into question certain aspects
of the (neo-)Darwinian theory of the evolution through natural selection. 
Why should not all the possible directions, i.e. all the possible mutations,
be statistically generated at the same time? Why should then evolution not be
a ``continuum'' process? This has even induced to talk about ``ontogenesis''
for this kind of mutations, and mathematical models have been investigated,
in order to explain this behaviour \cite{malasse2,malasse3,malasse5}.

We don't want here to delve into the problematic of the ``mechanics''
of this progressive contraction. We are more generally
interested in the biophysical dynamics of evolution, which seems
to occur through a sequence of steps forward and rests, and this
not only with regard to the
human species, but also more in general to the big Eras of life on 
the Earth.
In this note, we approach the problem from a point of view inspired
by our recent work in fundamental physics, \cite{spi}, 
where we consider the physics
arising in the framework of a non-perturbative string theory scenario.

\
\\

According to \cite{spi}, all fundamental mass scales $m_i$, 
as well as the couplings of elementary particles $\alpha_j$, are
expected to mainly run, during the cosmological evolution, 
as appropriate roots of the (inverse) age of the Universe.
Their dominant behaviour would therefore be of the type:
\be
m_i \, \sim \, {1 \over {\cal T}^{1 \over \gamma_i}} \, , ~~~
\alpha_j \, \sim \, {1 \over {\cal T}^{1 \over \gamma_j}} \, ,
~~~~ \gamma_i, \, \gamma_j \, > \, 1 \, ,  
\label{mialphaj}
\ee
where ${\cal T}$ is the age of the Universe, measured in reduced Planck units
(the units for which all the fundamental
constants, namely the Planck constant $\hbar$, the speed of light $c$,
and the Planck mass $1 / \sqrt{G_N}$ are set to 1), and $\gamma_i$, 
$\gamma_j$ are appropriate positive numbers, $\gamma_i, \gamma_j >1$. 
As a consequence of \ref{mialphaj}, in first approximation
also all atomic and molecular 
energy scales run, up to some normalization coefficients,
as appropriate powers of the (inverse) age of the Universe:
\be
E_p \, \sim \, {1 \over {\cal T}^{1 \over p}} \,  
+ \, {\cal O} \left( {1 \over {\cal T}^{1 \over q}}  \right) \, ,
~~~~ p > q > 1 \, .  
\label{ETpq}
\ee
Of course, in these units
the age of the Universe is an adimensional quantity, it
is a ``pure number'', so that
it must not surprise that an energy can correspond to any power of a time.
In order to obtain something with the dimension of an energy, 
the r.h.s. of equation~\ref{ETpq}
must be eventually multiplied by the Planck mass times the speed
of light to the square. 
At our present time, the rate of variation of 
couplings, masses, and energies, is very small, irrelevant for our
experience of every day. However, it becomes significant as seen on a 
cosmological scale. But its effect is appreciable also
at ``intermediate'' scales, such as those of the
evolution of life, where it can show out in ``fine-tuning'' effects.  
Among  these are precisely the cases of natural evolution we are going
to discuss.

Here we will discuss how the sequence of these evolutionary steps, 
as well as the relatively short duration
of the intervals of ``rapid'' progress of the evolution,  can be explained 
entirely within the laws of molecular physics and the Neo-Darwinian theory of
natural evolution.

\section{The evolution of Primates}
\label{prim}

\begin{figure}
\vspace{.5cm}
\centerline{
\epsfxsize=8cm
\epsfbox{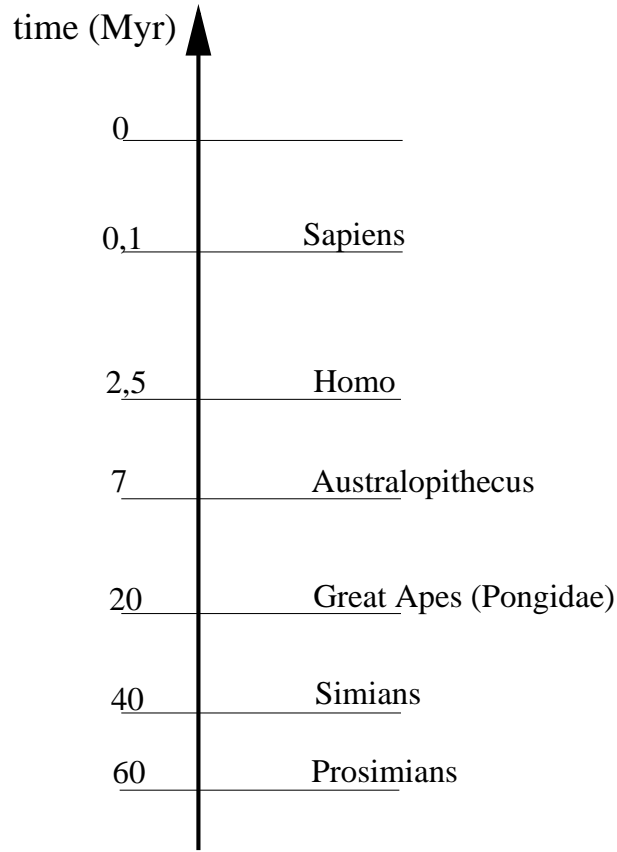}
}
\vspace{0.8cm}
\caption{The steps of increasing cranio-facial contraction of hominids, 
according to \cite{malasse1}.}     
\label{eras-h}
\vspace{.5cm}
\end{figure}

Let's consider first the example referring to the most recent
series of evolutionary mutations: the evolution of primates
along steps of increasing cranio-facial contraction, summarized in
figure~\ref{eras-h}. 
\noindent
It is clear that the duration of these periods increases
as we go back in time to earlier ages, although no simple mathematical 
relation seems to relate them. 
Once expressed in units of the age of the Universe,
the periods ${\cal T}_n $ of the primates-to-human history show a 
behaviour much less unfamiliar. Indeed, as we will discuss, they 
approximately arrange into a power series:
\be
{\cal T}_n \, \approx \, k \, n^q \, ,
\label{TiNq}
\ee 
for some positive numbers $k$ and $q$, $0 < q < 1$, and $n$ running on the
natural numbers. What produces this behaviour? 
The fact that mutations seem to occur
during a very short time, as compared to the duration of the ``stable'' 
phases, recalls the typical width of a resonance threshold in energy
absorption processes. In a quantum system,
energy levels are quantized and in general discrete; this is true at least
as long as we consider a bound system and its binding energies, 
a situation to which the DNA corresponds 
with good approximation.     
Mutagenesis is a process produced by a change in the DNA structure. 
At the molecular level, what happens is that,
as a consequence of the absorption of a certain amount of energy
(e.g. radiation of a certain frequency), protons and/or electrons ``jump'' to
different positions, and form new bonds.  
Let's consider to expose the DNA to a certain kind of radiation.
The energy that hits the probe is quantized, and is related to the frequency
$\nu$, or the wavelength $\lambda$, of the radiation, 
according to the Compton law:
\be
E_{\rm source} \, = \, h \nu  \, = \, {c \over \lambda} \, .
\label{compton}
\ee
Also the energy levels of the target molecule are expected to be
quantized. The typical energies of mutagenetic processes are the object 
of several investigations, 
based on approximations of the DNA sequence as a crystal, or in general
a system bound in a certain region 
\cite{chang-2000,frappat-1998-250,frappat-2000,diamant-2000-61}. In general, 
the absorption spectrum is discrete:
\be
E_{DNA} \, = \, \left\{  E(n)   \right\} \, , ~~~~ E(n) \, = \, k_n E_0 \, ,
\label{EDNA}
\ee
where $k_n$ is a certain coefficient and $n$ runs on (a subset
of) the natural numbers. The radiation energy \ref{compton}
can be absorbed by the DNA molecule, and produce a change in its structure,
only if it corresponds to one of the discrete levels of its spectrum. 
In this case, we have a resonance of the absorption probability:
\be
\left. E_{\rm source} \right|_{\rm res.} \, \cong \, E(n)_{\rm target} \, .
\label{EresEtarget}
\ee 
A series of evolutionary steps, such as those
of the progressive cranio-facial contraction,
corresponds to a specific change of the DNA structure, 
possibly induced by a change of one or more proton bonds, that could be
a transition of the kind considered in Ref.~\cite{chang-2000}, or something 
similar. Which molecular bonds do
correspond to a certain degree of contraction is not known.
However, it is not unreasonable to think that
the amount of contraction is related to the number of bonds 
which underwent an ``elementary'' transition in the DNA molecule. 
Let's make the \underline{hypothesis} that this is indeed the case.
A larger degree of mutation would then correspond to a larger number
of elementary transitions.
In order to induce one such change, an ``elementary step'' $A$,
the DNA molecule must absorb an energy:
\be
E_A \, = \, E(n_A) \, = \, k_{n_A} E_0 \, ,
\label{EnA}
\ee
for some quantum number $n = n_A$. Let's suppose
that this is precisely induced by the absorption of
energy coming from an external source of radiation.
In order to induce the evolutionary mutation under consideration,
we must therefore have:
\be
\left. E_{\rm source} \right|_{\rm res.} \, \cong \, E(n_A)_{\rm target} \, .
\label{EresEnA}
\ee 
A discrete series of resonance points along the time axis
is only possible if the two energy scales
run as independent functions of time. The amount of change at any such point
should be related to the time width of the resonance. 

\
\\

\noindent
As anticipated, let's suppose mutations are induced by radiation.
There are several candidates for a source
of radiation able to induce genetic mutations: 
the UV radiation, mostly coming from solar light;
the natural radioactivity, and cosmic rays.
However, X and cosmic rays are extremely energetic, and the mutations
they induce are in general not ``evolutionary'' but ``destructive''.
The radiation that in practice can induce molecular changes leading to
new forms of life, not just to the death of an organism, is the
ultra-violet, and perhaps even less energetic, radiation. 
Therefore, the energy spectrum of the source should 
basically be the one of the electronic transitions, giving rise
to the known atomic emission spectra (in the case of hydrogen, the
Lyman series etc...).

During the cosmological evolution, the spectrum and the amount of this
type of radiation have changed, according to
the evolution of the stars and in particular of the solar system.
However, for what matters our problem,
restricted to a very recent era of the evolution of the Universe,
it can be considered a sufficiently regular background \footnote{We refer
here to the frequencies of the spectrum, and in general
the cosmological running of the fundamental physical parameters.
We don't consider variations due, for instance, to the solar activity,
that don't affect such properties. We will comment about these effects 
in section~\ref{remarks}.}.
Were the energy levels of the source, and of the target DNA, constant (as
they are normally assumed to be), the mutation process would be progressive:
the elementary transition would be constantly related to a certain spectral
line, or a bunch of spectral lines. 
The rate of absorption would be proportional to the intensity of the source
(almost constant), leading to a 
statistically continuous increase of the number of changed bonds in the DNA
molecule. We would therefore observe a continuous evolution of primates.
In the framework of the physical scenario discussed in Ref.~\cite{spi},
both the emitted radiation, and the ground
energy scale of the DNA bonds, being functions of elementary energy
scales and couplings, have a dominant behaviour given as
in \ref{ETpq}. This means that, in first approximation,
they run as two independent powers of the age of the Universe:
\ba
E_{\rm source} & \approx & { k_{\rm s} \over {\cal T}^{p_{\rm s}} } \, ; 
\label{Es} \\
&& \nn \\
E_{\rm target} & \approx & { k_{\rm t} \over {\cal T}^{p_{\rm t}} } \, ,
\label{Et}
\ea
where $p_{\rm s}$, $p_{\rm t}$ are real numbers 
$0 < ( p_{\rm s}, p_{\rm t}) < 1$, and
$k_{\rm s}$, $k_{\rm t}$ are coefficients 
that collect the contribution of symmetry factors
and encode the dependence on the quantum numbers
labelling the energy levels. At a generic time ${\cal T}$,
the radiated energy doesn't correspond to any energy gap of the target.
Let's suppose that at a certain age ${\cal T}_i$ we have a resonance
with some spectral line of the source:
\be
E(n_A) \, \approx \, E_{\rm source} (n,m) \, , 
\label{Eresi}
\ee
where $(n,m)$ is a shorthand notation that indicates the quantum
numbers of the two energy levels involved in the transition
producing the radiation in the source. 
When \ref{Eresi} is satisfied, energy can be absorbed, making
possible for the system to undergo a class of genetic mutations, 
corresponding to new possible DNA molecular changes. Statistically,
in a short time, corresponding to the width of the resonance,
all possible mutations are tried out.
There is therefore not necessarily a unique kind of mutation.
The maximal transition probability is attained at the pick of the resonance. 
Out of this point, the
probability rapidly decays, at a speed depending on the characteristic
width of emission and absorption spectra.
In any case, after a ``short'' time, these transitions are no more possible
(i.e. they are extremely suppressed),
and the rate of the mutagenesis process drops down dramatically.
Natural selection will then decide which one(s) among all the mutations
will survive. The system will then ``stabilize'' until a new resonance
threshold opens up. 
Suppose this was a facial bone contraction enabling a larger brain volume;
we get a certain amount of contraction-inducing transitions 
(i.e. a certain amount of changed DNA bonds), depending on the width of the 
resonance window. Then the process stops till the new resonance.
This occurs when the same kind of molecular transitions are induced by the
next spectral line that turns out to meet the condition \ref{Eresi}. 
If a larger brain is a mutation favoured by natural selection also
at later times, then, at the
next resonance time, Nature will favour again the same kind of transition;
the suspended process of contraction will be resumed
and progress for another while, leading to the birth of species of
primates with a still larger brain.

We can give a rough estimate of the separation between subsequent
resonance times. First of all, let's see what is the order of magnitude we 
should expect for the exponents $p_{\rm s}$ and $p_{\rm t}$ of eqs.~\ref{Es} 
and \ref{Et}. For the emission scale, 
under the hypothesis of an atomic origin of the radiation, 
whatever is the source of radiation in first approximation
the atomic energy levels are given as some numbers multiplied by the Rydberg
constant $R$.
This is strictly true only in the simplest case, the hydrogen atom,
in which case the energy levels
are given by:
\be
E_{\rm source}(n,m) \, = \, h \nu \, = 
\, R \left( {1 \over m^2} - {1 \over n^2}  \right) \, ,
\label{vRyd}
\ee
where:
\be
R \, \approx \, R_{\infty} = m_e \alpha^2 / 4 \pi ~~ (\times \, c / \hbar) \, ,
\label{RRinfty}
\ee
where $m_e$ is the electron's mass and $\alpha$ the fine structure constant
(in our framework, neither of them is constant).
The highest energy, ultra-violet series, is obtained with $m = 1$
(Lyman series). 
More in general, the energy levels have more complicated expressions,
and, for heavy elements, with many electrons, one has to consider
also relativistic effects scaling as $m_e \alpha^4$.
However, as long as we are interested in a rough estimate, we will assume 
here that the energy levels of our source have an effective
approximate hydrogen-like spectrum. This hypothesis is on the other
hand supported by the consideration that 
hydrogen is the most common element in the Universe. We expect therefore that
the energy levels behave approximately as the Lyman series:
\be
E_{\rm source} \, \approx \, R_{\infty} \left(1 - {1 \over n^2}   \right) \, .
\label{sLyman}
\ee   
For the target DNA molecule, the energy levels of interest for us are those 
corresponding to a transition not among the positions of the electrons
but of the protons (see for instance Ref.~\cite{chang-2000,golo-2001}).
We don't know what is the dominant term in the typical energy scale
of mutagenetic transitions; therefore, here we leave open the
possibility that the fundamental, time-dependent part of the binding energies 
could even be more sensitive to the proton (and neutron) mass,
than to the electron's mass. Even in the lack of a precise knowledge
about the details of the DNA energies, let's suppose that their fundamental
scale is different from the one of the source of radiation.
In this case, there are two possibilities:
\begin{enumerate}
\item[1)]
The DNA fundamental scale runs slower, and therefore is above, the
scale of the source. In this case, since $1 / m^2 - 1/ n^2 < 1$, a resonance
of the source with the target, \ref{Eresi}, is only possible if:
\be
k_{n_A} \, < \, 1 \, ,
\label{knAvalue}
\ee
with $k_{n_A}$ as defined in \ref{EnA}.
This could be the case if we think that, having to deal with
energy levels related to bound states of protons, instead than of electrons,
for what concerns the transitions
of interest for us, the fundamental DNA
energy scale $E_0 = E_0^{\rm target}$, defined in \ref{EnA}, roughly
has a dominant behaviour analogous to the one of the atomic scale,
\ref{RRinfty}, but with the proton mass instead than the electron's mass. 
Namely, it could be something like
$E_0^{\rm target} \, \approx \, m_{\rm p} \alpha^2 $.
\item[2)]
The DNA fundamental scale runs faster, and therefore is below, the
scale of the source. This can be the case if 
the dependence on the coupling $\alpha$ is realized through
a higher power, something that would 
reduce the scale to lie below the scale \ref{RRinfty},
by ``eating'' the gain due to a
higher proton mass, or simply by suppressing by a higher amount the
electron's mass scale, so that:
\be
E_0^{\rm target} \, \approx \, m_{({\rm p}/e)} \alpha^{\beta} 
~ < \, E_{\rm source}\, , ~~~~~~~ \beta > 2 \, ,
\label{E0target2}
\ee  
and
\be
k_{n_A} \, > \, 1 \, .
\label{knAvalue2}
\ee
\end{enumerate}
According to \cite{spi}, both the electron mass and the
electric charge (the fine structure constant $\alpha$) run as positive roots
of the inverse of the age of the Universe. 
This means that the Rydberg constant too
scales as a certain root of the age of the Universe.
At sufficiently large times as compared to the Planck length (as is the case
of the evolution of life),
also the proton mass roughly scales as a root of the age of the Universe. 
With reference to equations~\ref{Es} and \ref{Et}, we can therefore
identify:
\ba
{1 \over {\cal T}^{p_{\rm s}}} & \sim & R_{\infty}\, = 
\, R_{\infty}({\cal T}) 
\, \equiv \, E^0_{\rm source}({\cal T})
\, ; \label{TpsR} \\
{1 \over {\cal T}^{p_{\rm t}}} & \sim & E^0_{\rm target}({\cal T}) \, .
\label{TptE0}
\ea
For the purpose of the present discussion, there is no fundamental
difference between case (1) or (2).
Important for our argument is just that we assume that the DNA ground energy
scale runs with time differently from the scale of the source. 
However, the choice of the one or the other of (1) and (2) implies a deep 
difference in the interpretation
when we consider larger time scales, as we will comment at the end of the
analysis. 
If we suppose here that the DNA fundamental scale runs
slower than (and therefore is above) the atomic scale of the source,
case (1), then:
\be
p_{\rm s} \, > \, p_{\rm t} \, .
\label{ps>pt}
\ee
If instead we suppose that the DNA fundamental scale
runs faster than (and therefore is below) the atomic scale of the source,
case (2), we have:
\be
p_{\rm s} \, < \, p_{\rm t} \, .
\label{ps<pt}
\ee
In both cases, the resonance condition \ref{Eresi} at a time
${\cal T}_i$ can be written as:
\be
{\cal T}_i^{p_{\rm s} - p_{\rm t}} \, \approx  \, k_{n_A} \, \times \, 
\left(1 - {1 \over n_i^2}   \right) \, .  
\label{TnRes}
\ee
In case (1) the series of ${\cal T}_i$ progresses toward higher $n$
(higher energy levels of the source);
in case (2) the series of ${\cal T}_i$ progresses toward lower $n$ (lower 
energy levels of the source).

We can work out what is the sequence of times between such resonances,
namely, the differences:
\be
{ {\cal T}_{i + 1} - {\cal T}_i   } \, , ~~
{ {\cal T}_{i + 2} - {\cal T}_{i+1}   } \, , ~ ~~ \ldots  ~~  ,
\label{Tii+1}
\ee 
by solving the equation~\ref{TnRes} for $n_i = n$, $ n_{i + 1} = n +1$, 
$n_{i+2} = n + 2, \ldots$ if $p_{\rm s} > p_{\rm t}$ (case 1),
and for $n_i = n$, $ n_{i + 1} = n - 1$, 
$n_{i+2} = n - 2, \ldots$ if $p_{\rm s} < p_{\rm t}$ (case 2). 
Let's introduce $q \equiv 1 / (p_{\rm t} - p_{\rm s})$. Clearly, $|q| > 1$; 
we can then write equation~\ref{TnRes} as:
\be
{\cal T}_i \, \approx \,  \left[ k_{n_A} \, \times \, 
\left(1 - {1 \over n_i^2}   \right) \right]^{q} \, ,
\label{Tqn}
\ee
where the choice of sign of the exponent, $\pm \, |q|$, depends on whether
the physical situation corresponds to case 1) or 2).

\begin{figure}
\vspace{.5cm}
\centerline{
\epsfxsize=10cm
\epsfbox{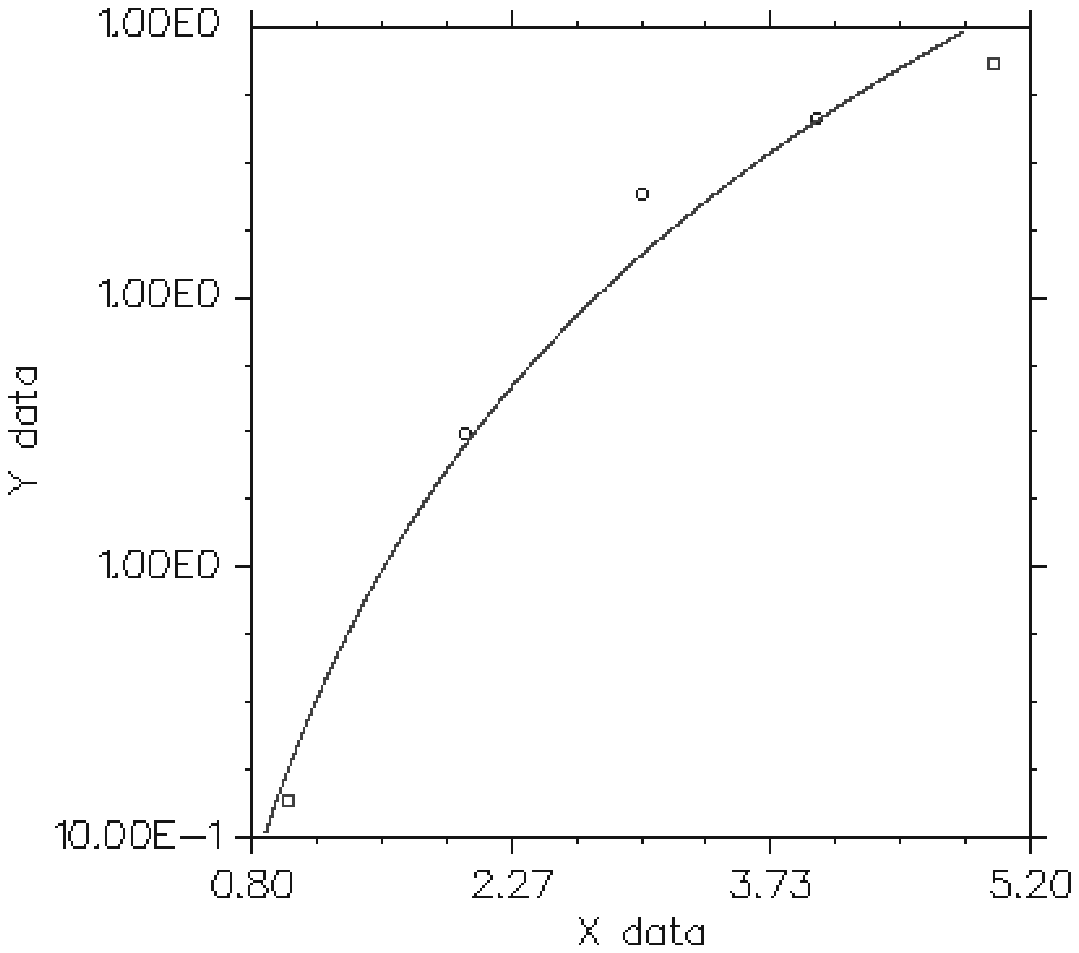}
}
\vspace{0.8cm}
\caption{interpolation of human evolution with the curve 
$y = a x^c$.}     
\label{interp-eras-no-shift}
\end{figure}

\begin{figure}
\vspace{.5cm}
\centerline{
\epsfxsize=10cm
\epsfbox{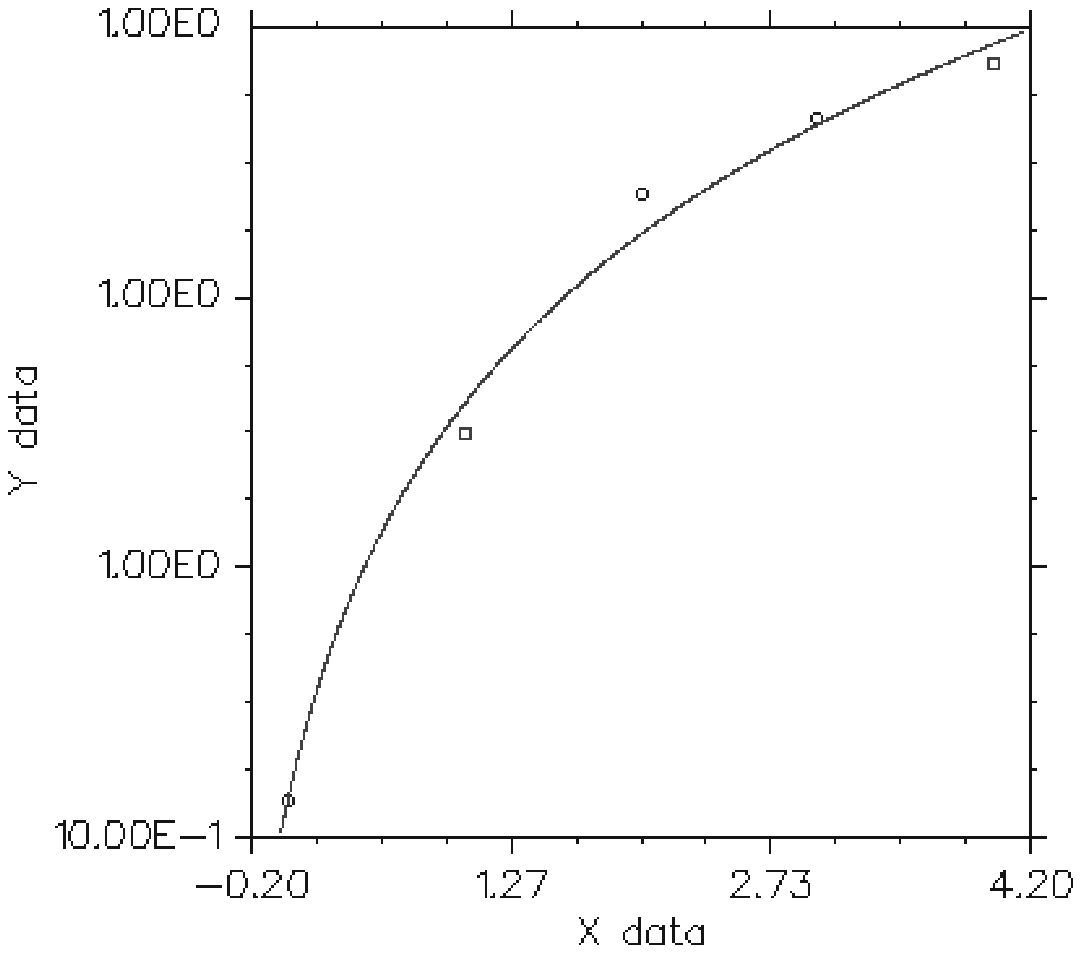}
}
\vspace{0.8cm}
\caption{interpolation of human evolution with the curve
$y = a (x-b)^c$.}     
\label{interp-eras-shift}
\end{figure}

In order to verify our hypothesis, we fit equation \ref{Tqn} over
five points in the history of the Universe, corresponding to the turning
periods in which mutagenesis has produced the 
evolution of the human species from the Australopithecus
to the Homo Sapiens, illustrated in figure~\ref{eras-h} of 
page~\pageref{eras-h}.
A first problem of such a numerical computation
is that the age of the Universe is only known with a rough approximation.
The common estimates range from 11,4 to 15 billion years.
As we discussed in Ref.~\cite{spi},
this value could be an over-estimate: within the framework of 
\cite{spi} everything seems to be consistent with a slightly shorter
age, of around 9,6 billion years. 
To be ``conservative'', we will assume an age of the Universe
of around 10 billion years. After all, we are here interested in just a
rough estimate: other, perhaps larger, inaccuracies could affect
our calculation. 
A major problem of this interpolation is however that 
the age of the universe, measured in reduced Planck units, whatever its value
precisely is, is an extremely huge number: $\sim {\cal O} (10^{60})$.
Generic curve-fitting algorithms are not able to deal with such
numbers, and try to find the best interpolation by reducing
the parameters to numbers of order 1. In order to get rid of big numbers and
constant parameters, we plot therefore the quantity:
\be
y(x) \, \equiv \, {{\cal T }_{\ell + x} \over {\cal T }_{\ell}} \, , 
\label{yxell}
\ee
for the five values from ``Simians'' to ``Sapiens'' as given in 
figure~\ref{eras-h}\footnote{We exclude the edge value corresponding to 
the prosimians, on which we will comment later.}. 
From expression~\ref{Tqn} we obtain:
\be
{{\cal T }_{i + N} \over {\cal T }_i} \, \cong \,
\left[ { 1 - {1 \over (n_i + ({\rm sgn} \,  q) N)^2} \over 
1 - {1 \over n_i^2}  } \right]^q \, .
\label{TnN/n}
\ee
For mass and energy scales ranging at present time from the meV to the keV
scale, the exponents $p_{\rm s}$ and $p_{\rm t}$ have typical values
in the range $\sim [{1 \over 2} \, $---$ \, {1 \over 2,4}]$. Therefore, 
$| q | \gg 1$. Limiting the analysis to the first
values of $N$, namely $N=1,2,3,4,5$, we can assume that
$N \ll n_i$. Under these conditions, expression \ref{TnN/n}
can be approximated by: 
\be
y(N) \,  \sim \,  N^c \, , ~~~~ N = 1,2,3 \ldots 
\label{yNc}
\ee 
for some constant $c$.
The small spacing of the periods, ${\cal T}_{i+1} - {\cal T}_i$,
as compared to the age of the Universe, tells us that $c \ll 1$. 
This approximation is valid as long as we can write:
\be
N ~ \approx ~ \left[ { 1 - {1 \over (n+ ({\rm sgn} \, q)
\tilde{N})^2} \over 
1 - {1 \over n^2}  } \right]^{q \over c} \, , ~~~~ 
\tilde{N} \equiv \pm \, (N-1) \, ,
\label{?approx}
\ee
where we have shifted the value of $N$ on the r.h.s. to $\tilde{N} = 
(N-1)$ in order to account for the fact that the point $N = 1$ of the 
interpolation corresponds to the point $\tilde{N} = 0$ on the r.h.s. 
Notice that, while the sequence of numbers on the l.h.s. increases
over the natural numbers, on the r.h.s. the sequence runs over the integers.
Namely, in the case the exponent $q < - 1$, the progression is toward 
decreasing energy levels of the source. This is obvious, because in this case
as time goes by the DNA scale becomes smaller and smaller as compared to
the scale of the source, and the resonance is realized with lower
energies of the source. 
For $n$ sufficiently large, $n > | \tilde{N}| $, 
we can expand the r.h.s. of \ref{?approx} as:
\be
\left[ { 1 - {1 \over (n+\tilde{N})^2} \over 
1 - {1 \over n^2}  } \right]^{q \over c} \, \approx \,
\left[ 1 \, \pm \, {2 | \tilde{N}| \over n^3}  \, +  \, 
{\cal O} \left( {1 \over n^2} \times 
\left( {\tilde{N} \over n }\right)^2 \right) 
\, \ldots \right]^{\pm {|q | \over c}} \, .
\label{rhsexpan}
\ee
By keeping just the first two terms of the expansion, we have a binomial
raised to the power $q/c$, and we obtain:
\be
N ~ \approx ~ 1 \, + \, \left( {q \over c}  \right)
{2 \tilde{N} \over n^3} \,  + \, \ldots
\, ,
\label{Nexpansion}
\ee
where the neglected terms receive a contribution from what we neglected in
\ref{rhsexpan}, of order:
\be
\sim ~ {\cal O}\left[ \left( { \tilde{N} \over n^2} \right)^{2} \right] \, ;
\label{O1}
\ee
and from the higher order terms in the binomial expansion:
\be
\sim ~ {\cal O}\left[ \left( { 2 \tilde{N} \over n^3} \right)^{2} \right] \, .
\label{O2}
\ee
The term $( q / c) 2 \tilde{N} / n^3$ in eq.~\ref{Nexpansion} is always
positive, because either is $q > 0$, with an increasing sequence of numbers in 
the atomic source, $\tilde{N} > 0$, or is $q < 0$, and the matching condition
is realized through a series of decreasing energies of the source,
$\tilde{N} < 0$. 
In either case, equation \ref{Nexpansion} is approximately solved by:
\be
n ~ \sim ~ \left( {2 | q | \over c} \right)^{1 / 3} \, .
\label{n3q/c}
\ee
Notice that this kind of approximation may work also
for atomic sequences other than the Lyman series.
For a generic $1 / m$ in expression~\ref{vRyd} 
we would obtain an expression analogous to~\ref{Nexpansion}, simply
with rescaled quantities: $n \to n/m$, $\tilde{N} \to \tilde{N}/m$,
resulting in a solution:
\be 
n ~ \sim ~ \left( {2 m^2 | q | \over c} \right)^{1 / 3} \, .
\label{nm3q/c}
\ee
Therefore, we don't really need to assume that the energies of the source
correspond to the Lyman series.
For what we have just discussed, it is reasonable to 
fit the ratios \ref{TnN/n}, referred to the five last steps 
of the evolution of primates, with the curve:
\be
y \, = \, a \, x^c \, .
\label{yxcurve}
\ee
Assuming an age of the universe of $\sim 10^{10}$ yr,
the values ${\cal T}_i$ can be approximated as:
\ba
{\cal T}_1 & \approx & 1,002 \times 10^{10} \, {\rm yr} \, ; \nn \\
{\cal T}_2 & \approx & 1,004 \times 10^{10} \, {\rm yr} \, ; \nn \\
{\cal T}_3 & \approx & 1,0053 \times 10^{10} \, {\rm yr} \, ; \nn \\
{\cal T}_4 & \approx & 1,00575 \times 10^{10} \, {\rm yr} \, ; \nn \\
{\cal T}_5 & \approx & 1,00599 \times 10^{10} \, {\rm yr} \, . \nn \\
\ea
By testing several interpolation options,
we have seen that it doesn't make
a big difference to fit the curve \ref{yxcurve} or to allow for a shift
of the $x$ value, namely the curve:
\be
y \, = \, a (x-b)^c \, .
\label{yxcurveb}
\ee
In any case, the computer solves the problem by finding
a very small exponent $c$, and a parameter $a$ of order 1.
The results are plotted in figures~\ref{interp-eras-no-shift}
and \ref{interp-eras-shift}. The curve fitting
gives in the case \ref{yxcurve}:
\ba
a & = & 1.0001526590176328 \, ; \nn \\
c & = & 2.5513644365246610 \times 10^{-3} \, ,
\label{ac-eras}
\ea
and, in the case of a shifted power, \ref{yxcurveb}:
\ba
a & = & 1.0015925425865762 \, ; \nn \\
b & = & 3.8904855104367164 \times 10^{-1} \, ; \label{abc-eras} \\
c & = & 1.6992267577200736 \times 10^{-3} \, . \nn
\ea 
We have tested the interpolation in several
ways, and seen that there can be a certain variation of these parameters,
according to the interpolation algorithm, 
and the preferences set for the interpolation.
Their values are therefore only roughly indicative. Within this 
approximation, the fit is also not so sensitive to a slight variation
of the value of the age of the Universe. 

\subsection{The step of Prosimians}
\label{pros}

At large $n$, the atomic energy levels get closer and closer
to each other (the atomic series converge to specific frequencies at the
limit $n \to \infty$). 
In the case (2) a larger atomic quantum number $n$ is found at earlier times.
In this case, by going toward the early steps of the sequence 
of resonances, the emitted energies accumulate to a ``continuum'', which
could appear as a wider width of a unique resonance
\footnote{This even though the ratio of the fundamental scales, 
i.e. the one of the source and the one of the DNA bonds under consideration, 
undergoes an accelerated increase
as time goes by, and therefore the time intervals become
shorter.}. Moreover,
in this regime the time intervals become approximately equal.
The condition:
\be
{\cal T}_{n+1} \, - \, {\cal T}_{n+1} \, \approx \, 
{\cal T}_{n+2} \, - \, {\cal T}_{n+1} \, ,
\label{Tn+2+1}
\ee
is in fact approximately equivalent to:
\be
\left( 1 \, - \, {1 \over (n + 1)^2}  \right)^p \, - \, 
\left( 1 \, - \, {1 \over n^2}  \right)^p 
\, \approx \,
\left( 1 \, - \, {1 \over (n + 2)^2}  \right)^p \, - \, 
\left( 1 \, - \, {1 \over (n+1)^2}  \right)^p \, ,
\label{n+2+1}
\ee
which is verified up to orders ${\cal O}\left[ {1 \over (n+1)^2} \right]$.
For lower quantum numbers, the approximation \ref{?approx} neglects
smaller terms (\ref{O1}, \ref{O2}), but as we approach a larger quantum number,
also the terms neglected in the ``linear'' approximation \ref{Tn+2+1}
are small enough. It could be that the starting point of the ``prosimians'' 
period in figure~\ref{eras-h}, page~\pageref{eras-h}, apparently falling out
of the sequence, precisely lies at the transition between the two kinds of
approximations we are making.
Moreover, we can expect that early times are estimated with a lower accuracy.
Does this time lie at the border of the series, 
so that earlier resonances simply ``accumulate'' within the time width
of a unique resonance, in such a way that this appears as the first, ``large''
resonance time of this kind of mutation?

In the scenario (1), at earlier times the resonance is realized
with lower energy levels of the source. 
In principle, at very earlier times, the mutation
could be induced by other atomic series, lower than the Lyman one.
Indeed, this series corresponds to the ultraviolet light only at present time,
and according to the scenario (1)
at earlier ages of the Universe its energies were higher as compared with
those of the DNA bonds. In this case, it could be that the step corresponding
to the beginning of prosimians corresponds to a value of $n$ at which
our approximation, valid for $n$ sufficiently large, starts to fail,
and perhaps corresponds to the end of a lower series. However,
a rough numerical check doesn't speak in favour of this hypothesis.
If also the beginning of the prosimians era has to be accounted within
this series of evolutionary steps, then the scenario (2) seems to be favoured.
In this case, as ${\cal T}$ increases, more kinds of molecular transitions
become possible, in correspondence to higher molecular levels
becoming accessible to a resonance with the highest atomic series.
On the other hand, the time windows allowed for a mutation become narrower 
and narrower, because the scale $1 / {\cal T}^{p_{\rm s}- p_{\rm t}}$
becomes smaller. Owing to the smaller time width, in the average
a smaller number of elementary transitions occur during a resonance. As a 
consequence, it becomes also smaller the average increase 
of the cranio-facial contraction.
We should therefore expect that the evolution of the
species tends to ``smooth down''
toward more frequent but less dramatic changes.

\section{The great Eras of life: the Paleozoic, Mesozoic and Cenozoic steps}
\label{greatEras}

\begin{figure}
\vspace{.5cm}
\centerline{
\epsfxsize=8cm
\epsfbox{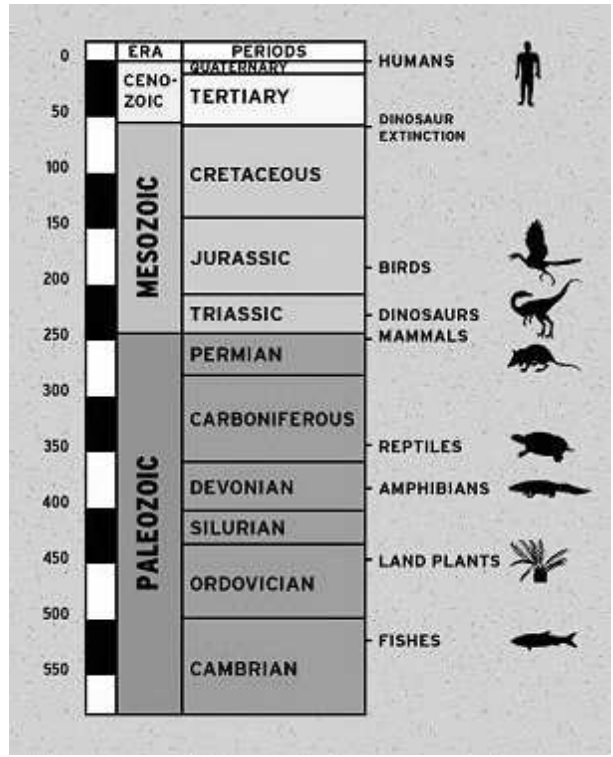}
}
\vspace{0.8cm}
\caption{The great Eras of the evolution of life.}     
\label{temps}
\vspace{1cm}
\end{figure}

The coincidence with a DNA absorption resonance
should in principle be at the ground of evolutionary processes that don't refer
only to the primates: it should work for any form of life. 
A problem is to identify which sets of mutations can be
grouped into classes corresponding to the same ``basic'' transition,
and therefore can be arranged along the same 
series of neighbouring resonances.
It is not hard to imagine that the evolutionary processes can be distinguished
into several classes, according to the kind of molecular transitions
they are controlled by.
For instance, by looking at figure~\ref{temps}, 
\noindent
one can figure out
that the big subdivision into Paleozoic, Mesozoic and Cenozoic Eras
of the natural evolution should not mix with the ``sub-eras'', the Periods
such as the Triassic, the Jurassic etc..., although these periods not 
necessarily fit into subclasses of the main class of transition.
This means that not necessarily ``Triassic, Jurassic and Cretaceous''
belong to the same main class, distinguished from the class formed
by the set ``Cambrian, Ordovician, Silurian, Devonian, Carboniferous,
Permian''.
The beginning of the first era, the Paleozoic Era, is 
the time when most of the major groups of animals first appear 
in the fossil record, and is sometimes called the "Cambrian Explosion",
because of the relatively short time over which this diversity of 
forms appeared. The Triassic-Permian extinction event too is something 
that took place in a relative short interval of time. Lastly, the end of the
Mesozoic era is characterized by the sudden disappearance of dinosaurs.
These facts strongly suggest that also the beginning
and the end of these eras were marked by a rapid evolution, as due to the
opening of new resonance thresholds allowing genetic mutation.
We may ask whether also these big eras of the 
evolution of life follow a power-law sequence.

Taking as ``resonance points'' the ages corresponding to the beginning of 
the Paleozoic era, the transition to the Mesozoic, and from Mesozoic to
Cenozoic, we obtain the following sequence: 
\ba
{\cal T}^{\prime}_1 & \approx & 1,0000 \times 10^{10} \, {\rm yr} \, ; \nn \\
{\cal T}^{\prime}_2 & \approx & 1,0350 \times 10^{10} \, {\rm yr} \, ; \nn \\
{\cal T}^{\prime}_3 & \approx & 1,0535 \times 10^{10} \, {\rm yr} \, , 
\label{Ttemps} 
\ea
where, for the sake of simplicity, we have rounded the basic time, at the
starting point of the Paleozoic era, to 10 billion years.
Actually, in figure~\ref{temps} there is a fourth age, our present time.
However, although, as discussed above, it seems that we are at turning point
of a new mutation process, this corresponds to a series of resonance
energies that, for what we have seen, it is safe to consider 
distinguished from the one we are considering now: 
it starts later, being entirely included within
the Cenozoic era. If the ages \ref{Ttemps} are going to belong to a 
sequence of resonances, this is quite probably another series, which
reaches the first resonance well before. 
All this to say that from our interpolation we exclude
our present time, that would correspond to:
\be
{\cal T}^{\prime}_4 \, \approx \, 1,0600 \times 10^{10} \, {\rm yr} \, .
\label{Tpresent}
\ee
By proceeding in the same way as in section~\ref{prim},
we plot the values $y(x)$. Unfortunately, we have only three data
for our interpolation. Nevertheless, the agreement of these data with
the fitting curve, as it can be seen from the
diagrams~\ref{interp-temps-3-no-shift} and~\ref{interp-temps-3-shift},  
is remarkable. 
The coefficient of the curve~\ref{yxcurve} are now:
\ba
a^{\prime} & = & 1.0003919688516729 \, ; \nn \\
c^{\prime} & = & 4.7632362251061745 \times 10^{-2} \, ,
\label{ac-paleo}
\ea
and, for the curve \ref{yxcurveb},
\ba
a^{\prime} & = & 1,0150031204496974 \, ; \nn \\
b^{\prime} 
& = & - 3,2534837743317041 \times 10^{-1} \, ; \label{abc-paleo-meso} \\
c^{\prime} & = & 3,7838527455527071E \times 10^{-2} \, , \nn
\ea 
where the $b^{\prime}$ coefficient comes with a negative sign because,
for computational reasons, we shifted the first $x$ value form 0 to 1.
Once shifted back, the correct $b$ coefficient is:
\be
\bar{b}^{\prime} = 1 - b^{\prime}= 0,67465162256682959 \, .
\label{bbar}
\ee 
As we already observed, the values of the interpolation coefficients
are only approximately indicative. A really significant output is   
on the other hand the fact that the coefficients
$c^{\prime}$ differ from the $c$ of section~\ref{prim}
by one order of magnitude.
This value is higher than the statistical uncertainty due to the artifacts
of the interpolation algorithm.
The difference between the two coefficients is therefore something real,
and signals that we are in the presence of absorption resonances
corresponding to a different series and power law.
This on the other hand is precisely what we should expect, on the base of 
the consideration that the genetic mutations are of another kind. 
In this case, they could correspond to different DNA transition energies. 
Indeed, the power-law behaviour \ref{yNc} is basically due to the
power-law scaling of the ratio of the basic scales
$E^0_{\rm source}/E^0_{\rm target}$, and the fact that within a certain 
range the quantum energy levels can be approximated by a simple harmonic 
oscillator-like expression $E(n) \approx  n E_0$. A quantum system in a box 
approximately correspond to a three-dimensional harmonic oscillator.
In the case of DNA, we can suppose that it roughly corresponds to a composite
system of many harmonic oscillators. In this way, at the first order
the coefficient $k_n$ in \ref{EDNA} should be given by:
\be
k_n \, \approx \, (n \, + \, n_0) k_0 \, ,
\label{kn0}
\ee  
where $k_0$ is a scaling factor and $n_0$ is the ground energy, a quantum 
Casimir effect that, if in the case of a one-dimensional harmonic oscillator
is $1/2$, in a complex system consisting of many harmonic oscillators
can be a much larger number. If this is the case, then, keeping fixed the
quantum numbers of the energy of the source, a power-law sequence
\ref{yNc} is obtained as long as we can approximate:
\be
\left( { n + \tilde{N} + n_0 \over n + n_0} \right)^{q \over c} ~
\approx ~ 1 \, + \, \left( {q \over c} \right) 
{\tilde{N} \over n + n_0} \, + \, 
{\cal O} \left( \tilde{N} \over n + n_0   \right)^2 \, ,
\label{nNn0}
\ee   
by retaining only the first two terms, and identifying this time:
\be
{q \over c} \, \sim \, n + n_0 \, , 
\label{q/c=n}
\ee
for some $n$. This is certainly possible, if the ground number $n_0$
is sufficiently large. In practice, the fact of having a sequence 
of the type~\ref{yNc} is related to the possibility of making
a linear approximation of
the spacing of the energy levels, either of the source or of the target,
or both of them,
into steps of equal separation, at fixed fundamental energy scale. 
Once the running of the latter is taken into account,
this translates into a series of the type~\ref{TiNq}.

Under these hypotheses,
our analysis tells us that also the three big eras of the evolution,
the Paleozoic, Mesozoic and Cenozoic, fit in a series of resonances.
According to these results,
we may ask whether the disappearance of dinosaurs, the event that
marks the end of the Mesozoic era, could be ascribed to the appearance
of more evolved competitors, perhaps coming from a mutation of
already existing species. It appears in fact the more and more clear that
their extinction, although it surely took place in a time interval very short 
as compared to the length of their period of existence, it has been a process
much longer than what we would have expected if it was
produced by some ``external'' catastrophic event, and perhaps better suits 
to a typical resonance width. 
We know that eventually mammals prevailed, although
they already existed well before; could it be that a slight mutation
finally gave them the necessary advantage to prevail over dinosaurs?

\begin{figure}
\vspace{.5cm}
\centerline{
\epsfxsize=10cm
\epsfbox{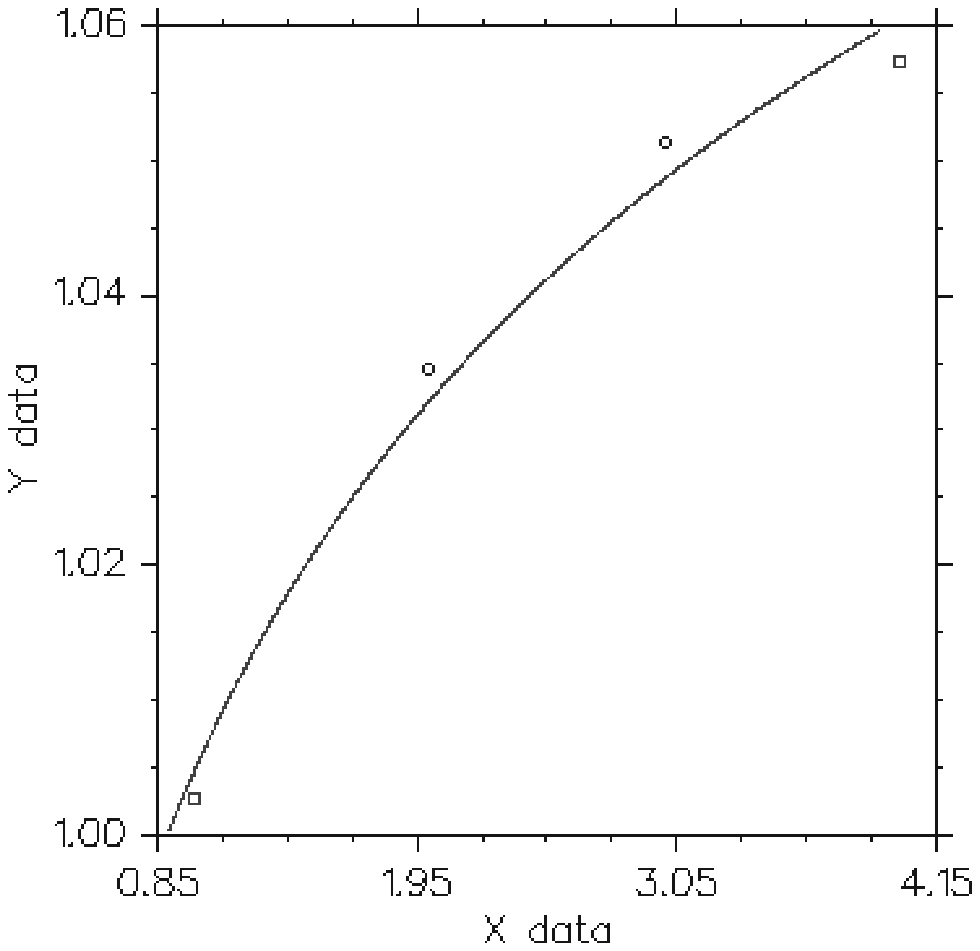}
}
\vspace{0.8cm}
\caption{interpolation of the eras of figure~\ref{temps} with the
curve $y = a x^c$ (4 values).}     
\label{interp-temps-4-no-shift}
\end{figure}

\begin{figure}
\vspace{.5cm}
\centerline{
\epsfxsize=10cm
\epsfbox{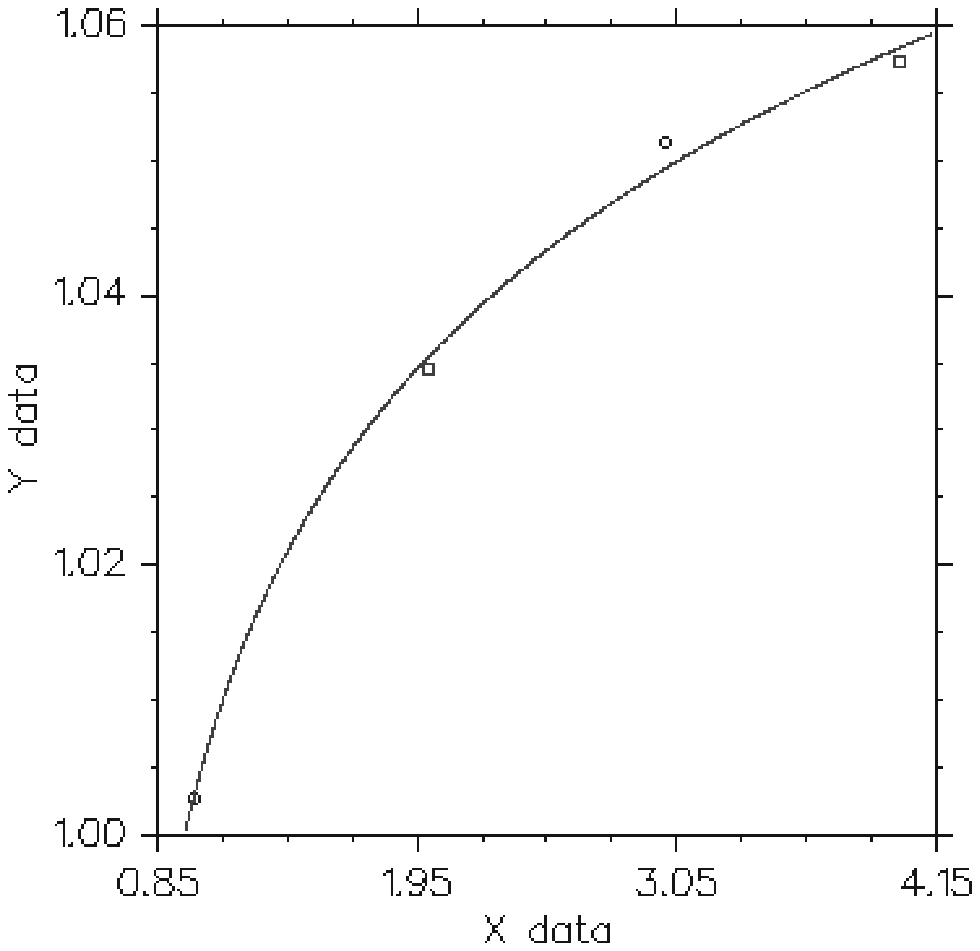}
}
\vspace{0.8cm}
\caption{interpolation of the eras of figure~\ref{temps} with the
curve $y = a (x-b)^c$ (4 values).}     
\label{interp-temps-4-shift}
\end{figure}

Plugging the coefficients \ref{ac-paleo} or \ref{abc-paleo-meso} 
in equation~\ref{yxcurve} (or \ref{yxcurveb} respectively),
we can also speculate about when should the Cenozoic era have an end.
Solving \ref{yxcurve} for $x = 4$ we obtain that this should correspond to
a time ${\cal T}_4$ such that:
\be
{{\cal T}_4 \over {\cal T}_1} \, \approx \, 1,069 \, ,
\label{end-no-shift}
\ee
and, in the case of the curve \ref{yxcurveb}:
\be
{{\cal T}_4 \over {\cal T}_1} \, \approx \, 1,073 \, .
\label{end-shift}
\ee
This means in around 9 million years, or 13 in the case
of the shifted-power curve \ref{yxcurveb}.
These predictions should be taken with a pinch of salt:
9, or 13 millions years, is a huge number as compared to the human history, 
but a little one if compared to the length of the Paleozoic and Mesozoic
eras: a difference of a dozen of millions in the estimate of the
length of the Cenozoic era would reflect in an error
of just a few percents in the estimate of the curve.
For instance, had we included in the interpolation also
our present time, given in \ref{Tpresent}, as the end point of an era,
we would have obtained the following fits:
\ba
a^{\prime \prime} & = & 1.0020109359413651 \, ; \nn \\
c^{\prime \prime} & = & 4.3037554472469014 \times 10^{-2} \, .
\label{ac-4temps}
\ea
and
\ba
a^{\prime \prime} & = & 1.0277362232536369 \, ; \nn \\
b^{\prime \prime} 
& = & - 6,4503233764454060 \times 10^{-1} \, ; \label{abc-4temps} \\
c^{\prime \prime} & = & 2.6484826008431502 \times 10^{-2} \, . \nn
\ea 
The plots are illustrated in figures~\ref{interp-temps-4-no-shift}
and~\ref{interp-temps-4-shift} , and show that
still the fits would be acceptable.
With these coefficients, the end point of the era starting
at our present time,
would be estimated to have its end in some 14 (resp. 15) millions years.

\section{Remarks}
\label{remarks}

\noindent
At this point, several considerations are in order:

\noindent
$\bullet$ Two different classes of the evolution,
namely the one of the big eras of life on the earth, and the one
of the primates, seem to arrange into sequences corresponding to DNA resonance 
energies. What distinguishes these two classes?
For what we have seen,
different series could be characterized by: 
\begin{enumerate}
\item A different ratio 
$k_{\rm s}/k_{\rm t}$, where $k_{\rm s}$ and $k_{\rm t}$ are given
in eqs.~\ref{Es} and \ref{Et}. This means
that the fundamental energy scales of both the source of radiation and the DNA
are the same, but the mutations are produced by transitions corresponding
to different energy levels of the same kind of source of radiation,
and/or different energy levels in the DNA bonds;
\item A differently running fundamental energy scale, 
either in the source (eq.~\ref{TpsR}), or in the target DNA 
(eq.~\ref{TptE0}), or in both of them. Since the time
dependence of these scales is a consequence of the time
dependence of the electron's mass and charge,
a different time scaling could be the consequence of a different
dependence of the energy levels on these quantities, as well as on
other time-dependent parameters, such as the proton mass.
This could be the case of mutations produced by another class
of DNA transitions.

\end{enumerate}

\noindent
Our approximation, and 
the small number of experimental data, together with their relative
inaccuracy, don't allow us to see finer differentiations and
discriminate between slightly different descriptions of the molecular
and atomic physics. A more accurate
analysis of the natural evolution could indeed provide some insight in the
structure of these energy levels, 
and open new perspectives to the investigation
of the DNA, providing more insight into its structure
and the dynamics of mutagenesis.
A small example of the possibilities offered by this method
is given by our discussion of the scenarios (1) and (2) of the evolution 
of the primates. In that case, a conclusion in favour of one of the
two possibilities (namely, the scenario 2) 
gives non trivial information about the DNA energy scales.

\begin{figure}
\vspace{.5cm}
\centerline{
\epsfxsize=10cm
\epsfbox{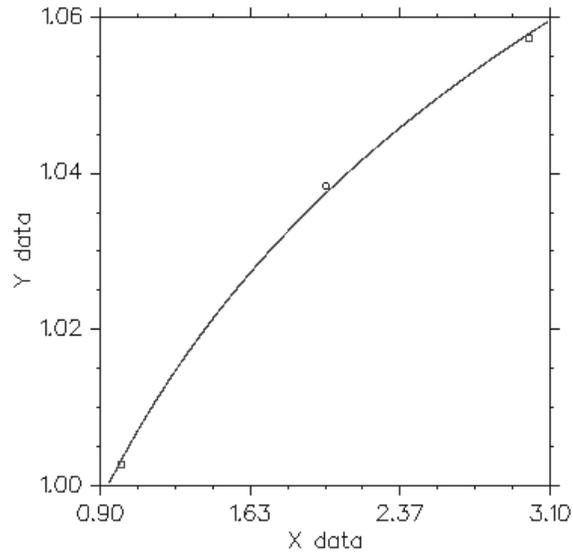}
}
\vspace{0.8cm}
\caption{interpolation of the duration of the 
eras of figure~\ref{temps} with the
curve $y = a x^c$ (3 values).}     
\label{interp-temps-3-no-shift}
\end{figure}

\begin{figure}
\vspace{.5cm}
\centerline{
\epsfxsize=10cm
\epsfbox{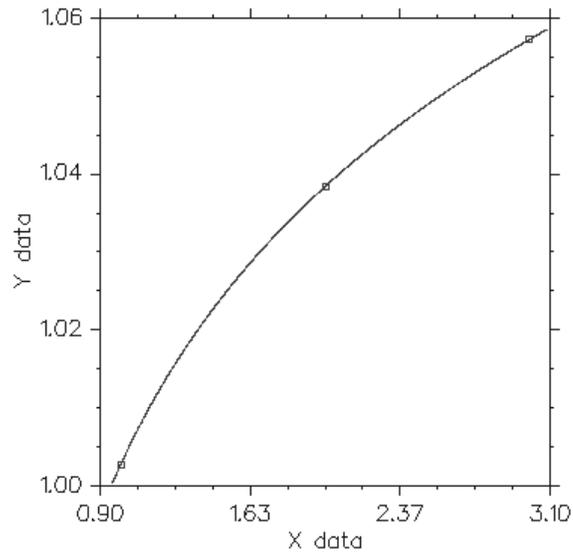}
}
\vspace{0.8cm}
\caption{interpolation of the eras of figure~\ref{temps} with the
curve $y = a (x-b)^c$ (3 values). 
Notice that here, for computational reasons, we
shifted the first $x$ value from 0 to 1 }     
\label{interp-temps-3-shift}
\end{figure}

At our present state of knowledge, we cannot decide out of any doubt
what distinguishes the sequence of the human evolution 
from the larger evolutionary
scale of the three main eras of figure~\ref{temps}.
In the case of the evolution of primates, we assumed that \emph{the same}
kind of molecular transition acts at any time there is a resonance condition.
The amount of progress in the evolution, 
according to \cite{malasse1} proportional to the 
amount of cranio-facial contraction, would then be proportional to the number
of occurred molecular transitions in the DNA. A priori it is
not clear whether also in the case of the sequence
of the big eras of figure~\ref{temps}, a unique kind of mutation is
at work during all the turning periods. Intuitively, we would say that 
in this case it is not, and the fact that the interpolation of these
periods with the power-law curve, figures~\ref{interp-temps-3-no-shift}
and \ref{interp-temps-3-shift}, gives an even better fit than in the
case of Primates, seems to be rather in favour of the 
interpretation that in this case the different turning times
correspond to different energy levels of the DNA, because in this case
the spacing is expected to be more regular. 
The question remains however open;
the seek for an answer could lead to a deeper understanding of the
mechanisms of DNA transitions and their relation to the evolution.

$\bullet$ Obviously, different molecular transitions lead to
different mutations. Therefore, the entire history of the evolution
cannot fit into a single series. However, in general not necessarily
all the steps of the evolution can be ordered into some series.
A simple look at eras, ages and periods, shows that there are many 
``irregular'' periods, which apparently cannot be arranged into any 
ordered sequence. Indeed, there can be a huge variety of
combinations of DNA and
source energy levels, leading to different mutations.
Owing to the superposition of different
mutations and different periods, the history of the evolution may not look
so easily well ordered. It remains however a key point that these transitions
occur at ``discrete'' points of the time axis, a feature that naturally fits
with our scenario of time-running energy scales.

$\bullet$ The time spread of a mutation period does not depend only on
the width of a resonance, but also on the fact that
natural radiation is not ``coherent'', it has a certain spread of frequencies.

$\bullet$ The main source of UV radiation coming to the earth is the sun.
Its activity is not constant; however, the solar phases involve the
amount of produced radiation, not its being in resonance or not.
As a consequence, under the hypothesis that the major cause of evolutionary
mutagenesis is the solar light, what we expect is that variations
of the solar activity
affect the evolution process only if they fall within the time window of some
resonance; in this case the mutation process can be accelerated
(or slowed down). 

$\bullet$ For simplicity, we did not consider mutagenesis of plants.
In principle, these too could (should?) follow similar laws, and perhaps 
the full story about evolution of species is the result of an 
interaction/interference of all these phenomena.

\
\\

All these considerations make only sense within the scenario
proposed in \cite{spi}, in which the energy scales depend on time.
Only in this case we obtain a discrete sequence of ``resonance'' periods.
Otherwise, the full spectrum of emission from natural sources, as well as
the complete spectrum of molecular energy levels, would be
fixed and constant all along the history. The conditions for a genetic
mutation would then be always the same, and 
mutations would be statistically generated without interruption.
A step-wise progress of the evolution would then require completely different
explanations.

We stress that,
when expressed in terms of the time separating these periods from
our present time, as in figures~\ref{eras-h} and~\ref{temps},
the power-law scaling, relation~\ref{TiNq},
is not explicit. The situation is similar to the one of the law of a perfect
gas, $PV = nR T$, in which the proportionality between
pressure/volume and the temperature is only unveiled when the latter
is expressed in terms of the absolute Kelvin scale.
Analogously, here in order to see the relation we must express the time 
periods in terms of the absolute age of the Universe.

Despite the caution one must have in considering curve fitting, 
and the large inaccuracy of data, 
made more dramatic by the small number of points among which to interpolate,
it remains a remarkable fact that, if expressed in terms of the
astronomical, ``absolute'' time scale, the main periods of the evolution 
of life seem to arrange into  series of steps corresponding to resonance
thresholds of typical molecular and atomic energy series.
This is by no means underestimating the role possibly
played by other factors, which
may act as ``disturbing'' agents, such as deep climatic changes
due to solar phases, meteorites, supernova-neutrino effects and so on.
And certainly, in the history of life many ``sub-periods'' seem to follow    
a more irregular path.
But certainly, it is intriguing to see that, perhaps, the main steps
are something ``regular'' and absolutely ``programmed''. Not by something
external to the rules of natural evolution and selection; simply, something
intrinsic of the fundamental laws of physics.

According to the scenario discussed in Ref.~\cite{spi}, 
the Universe is expected to evolve toward
more entropic configurations, in which the minimal energy step,
which is also the size of the ``unit cell'' of the phase space, decreases.
This agrees with the fact that the duration of the various phases
decreases, making the more and more frequent the transition points.
It however also means that the changes, the mutations, which are to be
expected, should become less dramatic: more frequent, but also in the average
smaller, steps.

\vspace{1.5cm}

\newpage

\providecommand{\href}[2]{#2}\begingroup\raggedright\endgroup

\newpage

\noindent

\noindent

\noindent

\noindent

\noindent

\noindent


\begin{thebibliography}{10}

\bibitem{spi}
A.~Gregori, \emph{An entropy-weighted sum over non-perturbative vacua,}
\href{http://www.arXiv.org/abs/arXiv:0705.1130 [hep-th]}{{\tt arXiv:0705.1130
  [hep-th]}}.
%%CITATION = ARXIV:0705.1130;%%.

\bibitem{malasse1}
A.~Dabrincourt-Malass\'{e}, \emph{Nouveau regard sur l'origine de l'homme,} La
  Recherche {\bf 286} (1996) 45--51.

\bibitem{malasse2}
A.~Dabrincourt-Malass\'{e}, \emph{L'hominisation et la th\'{e}orie des
  syst\`{e}mes dynamiques non lin\'{e}aires,} Revue de biologie
  math\'{e}matique {\bf 286} (1992) 117--119.

\bibitem{malasse3}
A.~Dabrincourt-Malass\'{e}, \emph{Continuity and discontinuity during
  modalities of hominization,} Quaternary International {\bf 19} (1993).

\bibitem{malasse5}
A.~Dabrincourt-Malass\'{e}, \emph{Modeling of cranio-facial architecture during
  ontogenesis and phylogenesis,} in {\em The Head-Neck sensory motor system}.
\newblock Oxford University Press, New-York-Oxford, 1992.

\bibitem{chang-2000}
C.-M. Chang, A.~H.~C. Neto, and A.~R. Bishop, \emph{Mutagenesis and Metallic
  DNA,} \href{http://www.arXiv.org/abs/cond-mat/0008166}{{\tt
  cond-mat/0008166}}.

\bibitem{frappat-1998-250}
L.~Frappat, A.~Sciarrino, and P.~Sorba, \emph{A crystal base for the genetic
  code,} Physics Letters A {\bf 250} (1998) 214.

\bibitem{frappat-2000}
L.~Frappat, A.~Sciarrino, and P.~Sorba, \emph{Crystalizing the genetic code,}
  2000.

\bibitem{diamant-2000-61}
H.~Diamant and D.~Andelman, \emph{Binding of molecules to DNA and other
  semiflexible polymers,} Physical Review E {\bf 61} (2000) 6740.

\bibitem{golo-2001}
V.~L. Golo and Y.~S. Volkov, \emph{Tautomeric Transitions in DNA,}
  \href{http://www.arXiv.org/abs/cond-mat/0110599}{{\tt cond-mat/0110599}}.

\end{thebibliography}
\end{document}